\documentclass[aps,pra,superscriptaddress,reprint]{revtex4-1}
\usepackage{amsmath}
\usepackage{amsthm}
\usepackage{amsfonts}
\usepackage{graphicx}
\usepackage{hyperref}
\newtheorem{thm}{Theorem}
\newtheorem{cor}[thm]{Corollary}

\theoremstyle{definition}

\theoremstyle{remark}

\begin{document}
\title{Universal locality of quantum thermal susceptibility}

\author{Giacomo De Palma}
\affiliation{QMATH, Department of Mathematical Sciences, University of Copenhagen, Universitetsparken 5, 2100 Copenhagen, Denmark}
\affiliation{NEST, Scuola Normale Superiore and Istituto Nanoscienze-CNR, I-56127 Pisa,
Italy.}
\affiliation{INFN, Pisa, Italy}

\author{Antonella De Pasquale}
\affiliation{NEST, Scuola Normale Superiore and Istituto Nanoscienze-CNR, I-56127 Pisa,
Italy.}

\author{Vittorio Giovannetti}
\affiliation{NEST, Scuola Normale Superiore and Istituto Nanoscienze-CNR, I-56127 Pisa,
Italy.}

\begin{abstract}
The ultimate precision of any measurement of the temperature of a quantum system is the inverse of the local quantum thermal susceptibility [De Pasquale et al., Nature Communications {\bf 7}, 12782 (2016)] of the subsystem with whom the thermometer interacts.
If this subsystem can be described with the canonical ensemble, such quantity reduces to the variance of the local Hamiltonian, that is proportional to the heat capacity of the subsystem.
However, the canonical ensemble might not apply in the presence of interactions between the subsystem and the rest of the system.
In this work we address this problem in the framework of locally interacting quantum systems.
We prove that the local quantum thermal susceptibility of any subsystem is close to the variance of its local Hamiltonian, provided the volume to surface ratio of the subsystem is much larger than the correlation length.
This result greatly simplifies the determination of the ultimate precision of any local estimate of the temperature, and rigorously determines the regime where interactions can affect this precision.
\end{abstract}

\maketitle
\section{Introduction}
Current technology permits the realization of extremely small thermometers \cite{pothier1997energy,peng2013nanowire,gao2002nanotechnology,giazotto2006opportunities}, conceived in order to carry out the challenging task of controlling the thermodynamical behaviour of physical systems at the spatial resolution of the micro- and nanometer length scales. In this regime the quantum correlations shared among the subcomponents of the system can play a non-trivial role in the measurement of the system temperature. Determining the ultimate precision of these  procedures represents a fundamental issue from many perspectives, such as the development and control of the products of the nanotechnologies~\cite{Schwab2000, Cahill2003, Linden2010,Mari2012} or for instance in the study of microorganisms \cite{cooney1969measurement}.
From a mathematical point of view, the action of a generic thermometer aimed to determine the inverse temperature $\beta$ of a global system $S$, can be modelled by a positive operator-valued measurement (POVM) \cite{breuer2007theory,nielsen2010quantum} on the subsystem $A$ with whom the thermometer directly interacts and which typically is a small fraction of $S$.

Let $\hat{\rho}(\beta)$ be the reduced state of the subsystem $A$ associated to the state of the global system $S$ at inverse temperature $\beta$.
The selected POVM will act on the state $\hat{\rho}(\beta)$ producing  an estimate $\tilde{\beta}$ for $\beta$ which represents the value returned by our thermometer.
The ultimate lower bound to the mean-square-error  of any such estimator is set by the quantum Cram\'er-Rao
bound~\cite{helstrom1967minimum,paris2009quantum,cramer2016mathematical} (see also \cite{ruppeiner1995riemannian,braunstein1994statistical,deffner2013thermodynamic})
\begin{equation}\label{CRB}
\Delta^2 \;\tilde{\beta}\geq1/F(\beta)\;,
\end{equation}
with $F(\beta)$ being the quantum Fisher information of the family of states $\left\{\hat{\rho}(\beta)\right\}_\beta$.
The functional $F(\beta)$  does not depend upon the specific choice of the detection process, instead it characterizes how  small variations in the inverse temperature $\beta$ of the global system $S$ influence the local state
of $A$: accordingly in Ref.~\cite{de2015local} it was used to gauge  the local thermal susceptibility of composite systems, large values of $F(\beta)$ corresponding to models where  the equilibrium temperature of $S$ is better perceived by its components.

If the subsystem $A$ can be described with the canonical ensemble, i.e.
\begin{equation}\label{eq:canA}
\hat{\rho}(\beta) = e^{-\beta\hat{H}_A}\left/\mathrm{Tr}_Ae^{-\beta\hat{H}_A}\right.\;,
\end{equation}
where $\hat{H}_A$ is the local Hamiltonian, the local quantum thermal susceptibility of the subsystem $A$ coincides with the variance of its local Hamiltonian, that is proportional to its heat capacity \cite{zanardi2007information,zanardi2008quantum,marzolino2013precision,marzolino2015erratum}:
\begin{equation}\label{FVar}
F(\beta)= \mathrm{Tr}_A\left[\hat{H}_A^2\,\hat{\rho}(\beta)\right] - \left(\mathrm{Tr}_A\left[\hat{H}_A\,\hat{\rho}(\beta)\right]\right)^2 \;,
\end{equation}
and we recover the well-known result of classical thermodynamics \cite{plischke2006equilibrium,kittel1980thermal,landau2013statistical}.
However, the assumption \eqref{eq:canA} is not always justified in  presence of interactions between the subsystem $A$ and the rest of the system \cite{gogolin2015equilibration}.

In this paper we address this problem in the framework of locally interacting quantum systems \cite{gogolin2015equilibration}, and we prove that \eqref{FVar} still holds in the high temperature regime for any subsystem $A$ whose volume to surface ratio is much larger than the correlation length (see Corollary \ref{cormain} below).
Locally interacting quantum systems constitute a very general framework that encompasses all the fundamental spin models, such as the Ising model \cite{ising1925beitrag}, the Heisenberg model \cite{bethe1931theorie}, the Potts model \cite{wu1982potts}, the $n$-vector model \cite{stanley1968dependence}, the Hubbard model \cite{hubbard1963electron} and the Majumdar-Ghosh model \cite{majumdar1969nextI,majumdar1969nextII}.
Our finding rigorously proves that when the correlation length is larger than the volume to surface ratio of the subsystem, local interactions do not spoil \eqref{FVar} and hence they do not significantly affect the precision of local measurements of the temperature.
In this case, the local quantum thermal susceptibility is determined by the expectation values of local operators (the local Hamiltonian and its square), and is therefore a local quantity itself.
This locality is universal since the volume to surface ratio of the subsystem does not depend on the microscopic details nor of the subsystem nor of the Hamiltonian.

\section{The setup}
We work in the framework of locally interacting many-body quantum systems, extensively discussed in the review \cite{gogolin2015equilibration}.
We hence consider a lattice characterized by a finite set $\mathcal{V}$ of sites (the ``quantum particles'' of the model).
The global Hilbert space $\mathcal{H}$ is the tensor product of the Hilbert spaces $\mathcal{H}_{\{x\}}$ associated to each site: $\mathcal{H}=\bigotimes_{x\in\mathcal{V}}\mathcal{H}_{\{x\}}$. The edges of the lattice, which represent the interactions among the particles, are  identified with  the subsystems $X\subset\mathcal{V}$ that are the support of one term $\hat{H}_X$
 of the system Hamiltonian
\begin{equation}\label{defH}
\hat{H}=\sum_{X\in\mathcal{E}}\hat{H}_X
\end{equation}
(the support of an operator $\hat{O}$ is defined as the smallest subset $Y\subset \mathcal{V}$ such that $\hat{O}$ acts like the identity outside $Y$, i.e. $\hat{O}=\hat{O}_Y\otimes\hat{\mathbb{I}}_{\mathcal{V}\setminus Y}$).
In Eq.~\eqref{defH}  the symbol $\mathcal{E}$ indicates the set formed by all of edges that compose the lattice which fully determines the geometry of the model.   In particular
the distance $d(x,y)$ between two sites $x,y\in\mathcal{V}$  is  the length $d$ of the shortest sequence of elements  $X_1,\ldots,X_d\in\mathcal{E}$ such that
$x\in X_1$, $y\in X_d$, and $X_i\cap X_{i+1}\neq\emptyset$ for $i=1,\ldots,d-1$.
Consequently  the distance between a site $x\in\mathcal{V}$ and a subsystem $Y\subset\mathcal{V}$ and the distance between two subsystems $X,Y\subset\mathcal{V}$ are given by
\begin{equation}\label{defd}
d(x,Y):=\min_{y\in Y}d(x,y)\;,\quad d(X,Y):=\min_{x\in X}d(x,Y)\;,
\end{equation}
respectively. Another important notion is   the boundary edge set $X_\partial$  of a subsystem $X\subset\mathcal{V}$: this
is the subset of $\mathcal{E}$ formed by the  (boundary) edges which overlap with both $X$ and its complement, i.e.
$X_\partial:=\left\{Y\in\mathcal{E}:Y\cap X\neq\emptyset,\;Y\cap\left(\mathcal{V}\setminus X\right)\neq\emptyset\right\}$.

The key property of locally interacting many-body quantum systems described by an Hamiltonian of the form~\eqref{defH}
is that they admit a critical  temperature $T^*$ above which  the correlation between any two observables decays exponentially with the distance between their supports~\cite{gogolin2015equilibration,kliesch2014locality}. Namely,
for any $|\beta|<\beta^*=1/(K_bT^*)$, $K_b$ being the Boltzmann constant, there exists a correlation length $\xi(\beta)>0$ such that for any two subsystems $X,Y\subset\mathcal{V}$ and any two observables $\hat{O}_X$, $\hat{O}_Y$ with support in $X$ and $Y$, respectively, we have
\begin{align}\label{eqcluster}
&\left|\langle\hat{O}_X\;\hat{O}_Y\rangle_\beta-\langle\hat{O}_X\rangle_\beta\langle\hat{O}_Y\rangle_\beta\right|\nonumber\\
&\leq4\left(\xi(\beta)+1\right)\min\left(\left|X_\partial\right|,\left|Y_\partial\right|\right)\|\hat{O}_X\|_\infty\|\hat{O}_Y\|_\infty e^{-\frac{d(X,Y)}{\xi(\beta)}},
\end{align}
with $\left|Z\right|$ denoting the number of elements of the set $Z$.
\begin{figure}[t!]
\includegraphics[width=\columnwidth]{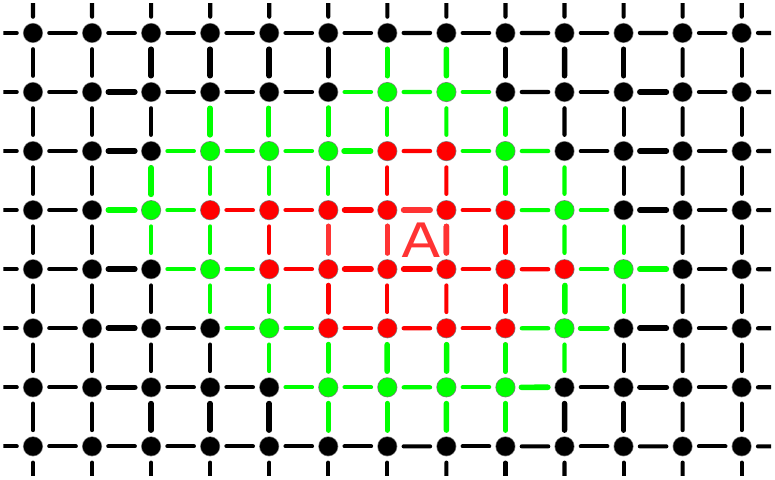}
\caption{A $2D$ square lattice with two-site first neighbours interactions.
In red the sites of $A$ and the edges of $\mathcal{A}$.
In green the sites $x$ with $d(x,A)=1$ and the edges of $\mathcal{C}_1$.
The boundary edges of $A$ are the green edges connecting a red with a green site.
In black all the other sites and edges. Here $N_\partial =6$. } \label{figC}
\end{figure}
In what follows we shall exploit the above inequality by focusing on models
 where the number of edges at distance $n\in\mathbb{N}$ from a given site $x\in\mathcal{V}$ increases at most quadratically with $n$, i.e. there exists a constant $M>0$ such that
\begin{equation}\label{surf}
\left|\left\{Y\in\mathcal{E}:d(x,Y)=n\right\}\right|\leq M\left(n+1\right)^2\;,
\end{equation}
for any $x\in\mathcal{V}$ and any $n\in\mathbb{N}$.
For a $D$-dimensional lattice, the left-hand side of \eqref{surf} scales as the surface of the sphere of radius $n$, i.e. as $n^{D-1}$.
Equation~\eqref{surf} reproduces the scaling for $D=3$.
For $D<3$, \eqref{surf} still holds since the actual scaling is tighter.
The case $D>3$ is unphysical.
However, replacing $(n+1)^2$ with $(n+1)^{D-1}$ in \eqref{surf} would only add an awkward $D$-dependent prefactor to our results.

\section{Universal locality}
In the presence of interactions the reduced density matrix of the subsystem $A$ might not be the Gibbs state \eqref{eq:canA}.
Quantum statistical mechanics~\cite{huang2009introduction,gogolin2015equilibration} assumes that this reduced density matrix is the partial trace of the Gibbs state of the global system:
\begin{equation}
\hat{\rho}(\beta)=\mathrm{Tr}_B\hat{\omega}(\beta)\;, \label{defrho}
\end{equation}
where $B$ is the remaining part of the system,
\begin{equation}\label{defomega}
\hat{\omega}(\beta)= e^{-\beta\hat{H}}\left/\mathrm{Tr}\,e^{-\beta\hat{H}}\right.\;,
\end{equation}
and $\hat{H}$ is the global Hamiltonian.
We stress that the assumption \eqref{defrho} does not require the global system to be in a Gibbs state.
It holds for almost all global pure states chosen from a given energy shell \cite{goldstein2006canonical,popescu2006entanglement}, and it is always satisfied for almost any time if the global Hamiltonian satisfies the Eigenstate Thermalization Hypothesis  \cite{deutsch1991quantum,srednicki1994chaos,tasaki1998quantum,calabrese2006time,cazalilla2006effect,rigol2007relaxation,reimann2007typicality,cramer2008exact,rigol2008thermalization,reimann2008foundation,linden2009quantum,rigol2009breakdown,rigol2012alternatives,reimann2010canonical,cho2010emergence,gogolin2011absence,riera2012thermalization,mueller2013thermalization,gogolin2015equilibration,polkovnikov2011colloquium,cazalilla2011one,bloch2008many,eisert2015quantum,deffner2015ten,jarzynski2015diverse,ponte2015many,steinigeweg2014pushing,genway2013dynamics,caux2013time,cassidy2011generalized,de2015necessity} (see also \cite{deffner2016foundations}).

We then suppose that the locally interacting many-body quantum system characterized by the Hamiltonian~\eqref{defH}
is initialized in the Gibbs state \eqref{defomega} whose inverse temperature $\beta$ we aim to recover via local measurements performed on
a subset  $A\subset\mathcal{V}$ of the sites. Let then $B=\mathcal{V}\setminus A$ be the complement of $A$ to $\mathcal{V}$,
$\mathcal{A}:=\left\{X\in\mathcal{E}:X\subset A\right\}$ the set of the edges contained into $A$, and
$\hat{H}_A:=\sum_{X\in\mathcal{A}}\hat{H}_X$, the local Hamiltonian of  $A$, i.e. the contribution to $H$ which act locally on $A$.
For any $R\in\mathbb{N}$, we define
\begin{equation}\label{defCR}
\mathcal{C}_R:=\left\{X\in\mathcal{E}\setminus\mathcal{A}:d(X,A)<R\right\}\;,
\end{equation}
the set of the edges not contained into $A$ and with distance from $A$ less than $R$, and
$\hat{H}_{\mathcal{C}_R}:=\sum_{X\in\mathcal{C}_R}\hat{H}_X$ the sum of the Hamiltonian terms in $\mathcal{C}_R$ (see \autoref{figC}).
For $R$ much smaller than the dimension of $A$, the edges of $\mathcal{C}_R$ are a thin layer of width $R$ that lies on the boundary of $A$.
Hence, we expect $\left|\mathcal{C}_R\right|$ to be proportional both to $R$ and to the surface of $A$, i.e.
\begin{equation}\label{NCR}
\left|\mathcal{C}_R\right|\propto R\left|A_\partial\right|\;.
\end{equation}

We can now state our main result: for $|\beta| < \beta^*$ the local quantum thermal susceptibility of the system $F(\beta)$
is close to the variance of the local Hamiltonian, namely
\begin{thm}\label{thmmain}
For any $R\geq2\xi(\beta)+1$
\begin{align}\label{eqmain}
&\left|\sqrt{F(\beta)}-\sqrt{\mathrm{Var}_\beta\hat{H}_A}\right| \leq \sqrt{\mathrm{Var}_\beta\hat{H}_{\mathcal{C}_R}}\nonumber\\
&+40JM|A|\sqrt{N_\partial}\left(\xi(\beta)+1\right)^\frac{3}{2}\times R^2e^{-R\left/2\xi(\beta)\right.}\;,
\end{align}
with $J:=\max_{X\in\mathcal{E}}\|\hat{H}_X\|_\infty$ being the interaction strength of  the system Hamiltonian $\hat{H}$ and
$N_\partial:=\max_{X\in\mathcal{E}}\left|X_\partial\right|$ being the maximum number of edges  sharing sites with a given one.
\begin{proof}
For any operator $\hat{O}$, we denote with $\left\langle\hat{O}\right\rangle_\beta := \mathrm{Tr}\left[\hat{O}\,\hat{\omega}(\beta)\right]$ its expectation value with respect to the global Gibbs state.
We remind that  the quantum Fisher information \cite{paris2009quantum} of a family of states $\left\{\hat{\rho}(\beta)\right\}_\beta$ can be expressed as
$F(\beta):=\|\frac{d}{d\beta}\hat{\rho}(\beta)\|_{\hat{\rho}(\beta)}^2$,
where given $\hat{\rho}$ a generic density matrix and $\hat{O}$ a generic operator, $\|\hat{O}\|_{\hat{\rho}}$ represents the Bures norm~\cite{bengtsson2007geometry},
\begin{equation}\label{Bdiag}
\|\hat{O}\|_{\hat{\rho}}=\sqrt{\sum_{i,\,j}\frac{2|\langle i|\hat{O}|j\rangle|^2}{p_i+p_j}}\;,
\end{equation}
with $\{ |i\rangle\}$ being the eigenvectors of $\hat{\rho}$ and  $p_i$ their corresponding eigenvalues.
Let then
\begin{equation}\label{defFR}
\mathcal{F}_R:=\left\{X\in\mathcal{E}:d(X,A)\geq R\right\}
\end{equation}
be the set of the edges with distance from $A$ at least $R$, such that $\mathcal{E}=\mathcal{A}\cup\mathcal{C}_R\cup\mathcal{F}_R$, and
$\hat{H}=\hat{H}_A+\hat{H}_{\mathcal{C}_R}+\sum_{X\in\mathcal{F}_R}\hat{H}_X$.
Given $\hat{\rho}(\beta)$ the density matrix of $A$ defined as in  \eqref{defrho},
we can then write
\begin{equation}
\frac{d}{d\beta}\hat{\rho}(\beta)=\hat{A}(\beta)+\hat{C}_R(\beta)+\sum_{X\in\mathcal{F}_R}\hat{B}_X(\beta)\;,
\end{equation}
where indicating with $\{\cdot,\cdot\}$ the anticommutator, we have
\begin{align}\label{defA}
\hat{A}(\beta) &=\{\langle\hat{H}_A\rangle_\beta-
\hat{H}_A,\;\hat{\rho}(\beta)
\}/2\;,\\
\hat{C}_R(\beta) &= \mathrm{Tr}_B [\{\langle\hat{H}_{\mathcal{C}_R}\rangle_\beta-\hat{H}_{\mathcal{C}_R},\;\hat{\omega}(\beta)\}]/2\;,
\end{align}
and for any $X\in\mathcal{F}_R$
\begin{equation}\label{defHX}
\hat{B}_X(\beta)  =\mathrm{Tr}_B[\{\langle\hat{H}_X \rangle_\beta -\hat{H}_X,\;\hat{\omega(\beta) }\}]/2\;.
\end{equation}
From the triangular inequality for the Bures norm it then follows that
\begin{align}\label{trin}
&\left|\|\frac{d}{d\beta}\hat{\rho}(\beta) \|_{\hat{\rho}(\beta) }-
\sqrt{\mathrm{Var}_\beta\;\hat{H}_A}
\right| \leq \|\hat{C}_R(\beta) \|_{\hat{\rho}(\beta) }\nonumber\\
&+\sum_{X\in\mathcal{F}_R}\|\hat{B}_X(\beta) \|_{\hat{\rho}(\beta) }\;,
\end{align}
where we used the fact that
$\|\hat{A}(\beta)\|_{\hat{\rho}(\beta)}=
\sqrt{\mathrm{Var}_\beta\;\hat{H}_A}$.
Next we invoke the fact that the Bures metric is contractive with respect to completely-positive trace-preserving (CPTP) maps \cite{nielsen2010quantum}. Accordingly we can bound the first term on the right-hand side of
~\eqref{trin} as follows
\begin{align}\label{FC}
\|\hat{C}_R(\beta) \|_{\hat{\rho}(\beta)}^2 &\leq \|\frac{1}{2}\{\langle\hat{H}_{\mathcal{C}_R}\rangle_\beta -\hat{H}_{\mathcal{C}_R},\hat{\omega}\}\|_{\hat{\omega}(\beta)}^2\nonumber\\
&=\mathrm{Var}_\beta\hat{H}_{\mathcal{C}_R(\beta)}\;.
\end{align}
For the second term instead we use the fact that for all operators $\hat{O}$ and states $\hat{\rho}$ one has
$\|\hat{O}\|_{\hat{\rho}}^2\leq \mathrm{Tr}[\hat{\rho}^{-\frac{1}{2}}\;\hat{O}\;\hat{\rho}^{-\frac{1}{2}}\;\hat{O}]$  which follows from \eqref{Bdiag} since $p_i+p_j\geq2\sqrt{p_ip_j}$ for any $p_i,p_j\geq0$.
Accordingly for any $X\in\mathcal{F}_R$ we can write
\begin{align}\label{ineq1}
&\|\hat{B}_X(\beta) \|_{\hat{\rho}(\beta)}^2 \leq \mathrm{Tr}\left[\hat{\rho}(\beta)^{-\frac{1}{2}}\;\hat{B}_X(\beta)\;\hat{\rho}(\beta)^{-\frac{1}{2}}\;\hat{B}_X(\beta)\right]\nonumber\\
&= \left\langle\hat{\rho}(\beta)^{-\frac{1}{2}}\;\hat{B}_X(\beta)\;\hat{\rho}(\beta)^{-\frac{1}{2}} (\langle\hat{H}_X\rangle_\beta- \hat{H}_X)\right\rangle_\beta \nonumber\\
&\leq 8\left(\xi(\beta) +1\right)N_\partial\,J \|\hat{\rho}(\beta)^{-\frac{1}{2}}\hat{B}_X\hat{\rho}(\beta)^{-\frac{1}{2}}\|_\infty e^{-d(X,A)/\xi(\beta)}\;,
\end{align}
where in  the second step we have used \eqref{defHX} and
in the last step we have applied \eqref{eqcluster} to the observables $\hat{O}_A=\hat{\rho}(\beta)^{-\frac{1}{2}}\hat{B}_X(\beta)\hat{\rho}(\beta)^{-\frac{1}{2}}$ and $\hat{O}_X=\langle\hat{H}_X\rangle_\beta- \hat{H}_X$ supported into $A$ and $X$, respectively,  bounding  $\left|X_\partial\right|$ with $N_\partial$ and using the inequality
\begin{equation}\label{eqHX}
\|\langle\hat{H}_X \rangle_\beta-\hat{H}_X\|_\infty\leq 2J\;.
\end{equation}
Now we notice that for any vector $|\psi\rangle$ in the Hilbert space of $A$ we have
\begin{align}
&\left|\langle\psi|\hat{\rho}(\beta)^{-\frac{1}{2}}\;\hat{B}_X(\beta)\;\hat{\rho}(\beta)^{-\frac{1}{2}}|\psi\rangle\right|\nonumber\\
&=\left|\mathrm{Tr}_B\left[\left(\langle\hat{H}_X\rangle_\beta-\hat{H}_X\right)\langle\psi|\hat{\rho}(\beta)^{-\frac{1}{2}}\;\hat{\omega}(\beta)\;\hat{\rho}(\beta)^{-\frac{1}{2}}|\psi \rangle\right]\right|\nonumber\\
&\leq2J\left\|\langle\psi|\hat{\rho}(\beta)^{-\frac{1}{2}}\;\hat{\omega}(\beta)\;\hat{\rho}(\beta)^{-\frac{1}{2}}|\psi\rangle\right\|_1=2J\langle\psi|\psi\rangle\;,
\end{align}
where $\hat{H}_X$ is meant as an operator on the Hilbert space of $B$, and where we have used \eqref{defHX} and \eqref{eqHX}.
Then
$\|\hat{\rho}(\beta)^{-\frac{1}{2}}\hat{B}_X(\beta)\hat{\rho}(\beta)^{-\frac{1}{2}}\|_\infty\leq2J$,
and \eqref{ineq1} implies
\begin{equation}
\|\hat{B}_X(\beta)\|_{\hat{\rho}}\leq 4J\sqrt{N_\partial\left(\xi(\beta)+1\right)}\;e^{-\left.d(X,A)\right/2\xi(\beta)}\;.
\end{equation}
Recalling \eqref{defFR} we can hence write
\begin{align}\label{FF}
&\sum_{X\in\mathcal{F}_R}\|\hat{B}_X(\beta)\|_{\hat{\rho}(\beta)}\leq 4J\sqrt{N_\partial\left(\xi(\beta)+1\right)}\nonumber\\
&\phantom{\leq}\times \sum_{n=R}^\infty\left|\left\{X\in\mathcal{E}:d(X,A)=n\right\}\right|e^{-n/2\xi(\beta)}\nonumber\\
&\leq 40JM\left|A\right|\sqrt{N_\partial}\left(\xi(\beta)+1\right)^\frac{3}{2}R^2\;e^{-\left.R\right/2\xi(\beta)}\;,
\end{align}
where in the last step we used the hypothesis~\eqref{surf}, stating that
the number of edges at distance $n\in\mathbb{N}$ from  $A$ increases at most quadratically in $n$, i.e.
\begin{equation}\label{assM}
\left|\left\{X\in\mathcal{E}:d(X,A)=n\right\}\right|\leq M\left|A\right|\left(n+1\right)^2\;,
\end{equation}
(indeed  from \eqref{defd}, if $d(X,A)=n$ there exists at least one site $x\in A$ with $d(x,A)=n$),
and  the inequality $\sum_{n=R}^\infty\left(n+1\right)^2e^{-n/2\xi}\leq10R^2\left(\xi+1\right)e^{-R/2\xi}$, which holds for any $\xi>0$ and $R\geq2\xi+1$.
The inequality~\eqref{eqmain} finally follows by replacing
\eqref{FC} and \eqref{FF} in \eqref{trin}.
\end{proof}
\end{thm}
The result of Theorem~\ref{thmmain}
can be made more explicit by bounding the term $\mathrm{Var}_\beta\hat{H}_{\mathcal{C}_R}$, for example  by showing that
the variance of the local Hamiltonian of a subsystem scales linearly with the size of the subsystem.
\begin{thm}[Variance of local Hamiltonian]\label{thmvar}
Let $\mathcal{G}\subset\mathcal{E}$ be a set of edges, and let
$\hat{H}_{\mathcal{G}}:=\sum_{X\in\mathcal{G}}\hat{H}_X$
be the sum of the Hamiltonian terms in $\mathcal{G}$.
Then, for any $|\beta|<\beta^*$ the variance of $\hat{H}_{\mathcal{G}}$ increases at most linearly with the size of $\mathcal{G}$, i.e.
\begin{equation} \label{TEO2}
\mathrm{Var}_\beta\;\hat{H}_{\mathcal{G}}\leq 8N_\partial\,J^2\left|\mathcal{G}\right|N\,M\left(\xi(\beta)+1\right)^4\;,
\end{equation}
where $N:=\max_{X\in\mathcal{E}}|X|$.
\begin{proof}
Reminding the defintion of $J$ and $N_\partial$ and  using  Eq.~\eqref{eqcluster} we can write
\begin{align}
&\mathrm{Var}_\beta \;\hat{H}_{\mathcal{G}} =\sum_{X,Y\in\mathcal{G}}(\langle\hat{H}_X\;\hat{H}_Y\rangle_\beta -\langle\hat{H}_X\rangle_\beta \langle\hat{H}_Y\rangle_\beta)\nonumber\\
&\leq 4\left(\xi(\beta)+1\right)N_\partial\,J^2\sum_{X,Y\in\mathcal{G}}e^{-\left.d(X,Y)\right/\xi(\beta)}\nonumber\\
&=4\left(\xi(\beta)+1\right)N_\partial\,J^2 \nonumber \\
&\phantom{=}\times \sum_{X\in\mathcal{G}}
 \sum_{n=0}^\infty\left|\left\{Y\in\mathcal{G}:d(X,Y)=n\right\}\right|e^{-n/\xi(\beta)}\;,
\end{align}
which yields \eqref{TEO2} by invoking the
assumption \eqref{surf} as in \eqref{assM} and  the inequality $\sum_{n=0}^\infty\left(n+1\right)^2e^{-n/\xi} \leq2\left(\xi+1\right)^3$.
\end{proof}
\end{thm}

\begin{cor}[Universal locality of quantum thermal susceptibility]\label{cormain}
For any $R\geq2\xi(\beta)+1$,
\begin{align}\label{eqcor}
&\left|\sqrt{F(\beta)}-\sqrt{\mathrm{Var}_\beta\hat{H}_A}\right|\nonumber\\
&\leq 2J\sqrt{2MNN_\partial}\left(\xi(\beta)+1\right)^2\times\sqrt{\left|\mathcal{C}_R\right|}\nonumber\\
&\phantom{\leq}+40JM|A|\sqrt{N_\partial}\left(\xi(\beta)+1\right)^\frac{3}{2}\times R^2e^{-R\left/2\xi(\beta)\right.}\;.
\end{align}
\begin{proof}
Follows from Theorems \ref{thmmain} and \ref{thmvar}.
\end{proof}
\end{cor}
Theorem \ref{thmmain} and Corollary \ref{cormain} imply that the local quantum thermal susceptibility of the subsystem $A$ is close to the variance of its local Hamiltonian $\hat{H}_A$ if we can find $R$ such that the right-hand side of \eqref{eqmain} or \eqref{eqcor} is much smaller than $\sqrt{\mathrm{Var}_\beta\hat{H}_A}$.
This is possible whenever
\begin{equation}\label{condR}
\left|\mathcal{C}_{\xi(\beta)}\right|\ll\left|\mathcal{A}\right|\;,
\end{equation}
i.e., recalling \eqref{NCR}, whenever the volume to surface ratio of the subsystem $A$ is much larger than the correlation length:
\begin{equation}\label{condA}
\xi(\beta)\left|A_\partial\right|\ll\left|\mathcal{A}\right|\;.
\end{equation}
Indeed, the second term in the right-hand side of \eqref{eqmain} or \eqref{eqcor} decreases exponentially as $R$ increases, and it is small whenever $R\gg\xi(\beta)$.
From Theorem \ref{thmvar}, we expect $\mathrm{Var}_\beta\hat{H}_A$ to scale linearly with $\left|\mathcal{A}\right|$.
The first term in the right-hand side of \eqref{eqcor} scales with $R$ as $\sqrt{\left|\mathcal{C}_R\right|}$.
Then, we expect it to be much smaller than $\sqrt{\mathrm{Var}_\beta\hat{H}_A}$ whenever $\left|\mathcal{C}_{R}\right|\ll\left|\mathcal{A}\right|$.
These conditions can be both satisfied iff \eqref{condR} holds.
On the contrary if the subsystem $A$ is smaller than the correlation length, conditions \eqref{condR} and \eqref{condA} do not hold.
In this case, we expect that the correlations between $A$ and the rest of the system will spoil \eqref{FVar}, and the local quantum thermal susceptibility will no more be close to the variance of the local Hamiltonian.
This always happens when the temperature approaches a critical point and the correlation length diverges, or when $A$ is made only by few sites.

\subsection{Example: 1D spin chain}
Let us conclude our analysis with an example. We consider a 1D spin chain with nearest neighbours interactions. In this case $M=2$, $N_\partial=2$ and $N=2$. To this class belongs for example the well-known 1D Ising model \cite{ising1925beitrag} with Hamiltonian $\hat{H}=\sum_{i}\left(\hat{\sigma}_i^x \,\hat{\sigma}_{i+1}^x + h \,\hat{\sigma}_i^z \right)$, where $h$ is the external magnetic field and $J=\sqrt{1+h^2}$.

Our choice for the subsystem $A$ is a set of consecutive sites.
We have with this choice $|\mathcal{C}_R|=2R$ for any $R\in\mathbb{N}$.
The bound of Corollary \ref{cormain} is then
\begin{align}\label{bound1DR}
&\left|\sqrt{\frac{F(\beta)}{\mathrm{Var}_\beta\hat{H}_A}}-1\right| \leq 8J\sqrt{2}\left(\xi+1\right)^2\nonumber\\
&\times\left(\sqrt{\frac{R}{\mathrm{Var}_\beta\hat{H}_A}}+\frac{10\left|A\right|R^2\,e^{-\frac{R}{2\xi}}}{\sqrt{\left(\xi+1\right)\mathrm{Var}_\beta\hat{H}_A}}\right)\;,
\end{align}
where the correlation length $\xi$ reads (\cite{gogolin2015equilibration}, Eq. (163))
\begin{equation}
\xi(\beta) = - \frac{1}{ \ln \left( \sqrt[3]{3}\, e^{2\beta J} ( e^{2\beta J} - 1 ) \right)}.
\end{equation}
The optimal bound is the minimum of the right-hand side of \eqref{bound1DR} over $R\ge2\xi+1$.
From the discussion in the previous Section, this minimum is small whenever $\mathrm{Var}_\beta\hat{H}_A\gg\xi$, i.e. whenever $|A|\gg\xi$ since $\mathrm{Var}_\beta\hat{H}_A$ is proportional to $|A|$.

\section{Conclusions}
We have proved that the local quantum thermal susceptibility of any subsystem of a locally interacting quantum system is close to the variance of its local Hamiltonian, provided the volume to surface ratio of the subsystem is much larger than the correlation length.
This result determines the ultimate precision of any local measurement of the temperature.
Moreover, it rigorously proves that the interactions between the subsystem and the remaining part of the system can affect this precision only when the correlation length becomes larger than the volume to surface ratio of the subsystem.
This always happens when the correlation length becomes larger than the subsystem, e.g. in the proximity of a critical point when the correlation length diverges, or for a subsystem made by only few sites.

\section*{Acknowledgements}
GdP acknowledges financial support from the European Research Council (ERC Grant Agreement no 337603), the Danish Council for Independent Research (Sapere Aude Grant No. 1323-00025B) and VILLUM FONDEN via the QMATH Centre of Excellence (Grant No. 10059).
This work was supported by the EU Collaborative Project TherMiQ (Grant agreement 618074), and by the EU project COST Action MP1209 Thermodynamics in the quantum regime.

\bibliography{biblio}
\bibliographystyle{apsrev4-1}
\end{document}